\documentclass{article}
\usepackage{spconf,amsmath,graphicx}
\usepackage{amsfonts}
\usepackage{epsfig,amssymb,latexsym,graphics,float}
\usepackage{setspace,subfig}
\usepackage{citesort}
\usepackage{amsopn}
\usepackage{amssymb,times}
\usepackage[ruled,vlined,linesnumbered]{algorithm2e}
\usepackage{ifpdf}

\begin{document}
\begin{onecolumn}
\thispagestyle{empty}
\vspace*{0.6cm}
\begin{center}

\vspace{-.2in}

{\LARGE Robust Particle Filter by Dynamic Averaging of Multiple Noise Models
}\\


{\large \vspace{1cm}
\vspace{0.3cm}

Bin Liu\footnote{Corresponding author.
E-mail: {\sf bins@ieee.org}.\\
Copyright 2017 IEEE. Published in the IEEE 2017 International Conference on Acoustics, Speech, and Signal Processing (ICASSP 2017), scheduled for 5-9 March 2017 in New Orleans, Louisiana, USA. Personal use of this material is permitted. However, permission to reprint/republish this material for advertising or promotional purposes or for creating new collective works for resale or redistribution to servers or lists, or to reuse any copyrighted component of this work in other works, must be obtained from the IEEE. Contact: Manager, Copyrights and Permissions / IEEE Service Center / 445 Hoes Lane / P.O. Box 1331 / Piscataway, NJ 08855-1331, USA. Telephone: + Intl. 908-562-3966.} \vspace{0.5cm}

{$~$School of Computer Science and Technology, \\Nanjing University of Posts and Telecommunications}\\
{$~$Nanjing, Jiangsu, 210023, China}

\vspace{0.5cm}

Manuscript Submitted --- Sep. 5th, 2016\\
This Manuscript has been accepted by ICASSP 2017 as an oral presentation paper\\
} 

\end{center}

\end{onecolumn}
\newpage
\begin{twocolumn}
\title{Robust Particle Filter by Dynamic Averaging of Multiple Noise Models}
\name{Bin~Liu$^{\star}$
\thanks{$^\star$Address correspondence to bins@ieee.org. This work
was partly supported by the National Natural Science Foundation
(NSF) of China (Nos. 61302158 and 61571238), the China Postdoctoral Science Foundation (Nos. 2015M580455 and 2016T90483), the NSF of Jiangsu Province (No. BK20130869), Scientific and Technological Support Project (Society) of Jiangsu Province (No. BE2016776).}}
\address{School of Computer Science and Technology, Nanjing University of \\
Posts and Telecommunications, Nanjing, 210023 China}
\maketitle
\begin{abstract}
State filtering is a key problem in many signal processing applications.
From a series of noisy measurement, one would like to estimate the state
of some dynamic system. Existing techniques usually adopt a Gaussian noise assumption
which may result in a major degradation in performance when the measurements are
with the presence of outliers. A robust algorithm immune to the presence of outliers is desirable.
To this end, a robust particle filter (PF) algorithm is proposed, in which the heavier tailed Student's t distributions
are employed together with the Gaussian distribution to model the measurement noise.
The effect of each model is automatically and dynamically adjusted via a Bayesian model averaging mechanism.
The validity of the proposed algorithm is evaluated by illustrative simulations.
\end{abstract}
\section{Introduction}\label{sec:intro}
This paper focuses on nonlinear state filtering, a key problem in many signal processing applications.
The aim here is to derive a novel particle filter (PF) algorithm that is robust towards outliers in the measurement noise.
We regard filtering with outliers as a model uncertainty problem, and address it using a multiple model strategy (MMS).
The MMS is a generic approach to handle model uncertainty problems. For example, in \cite{yi2016robust} and \cite{liu2011instantaneous}, the MMS is utilized to take account of the issue of measurement model uncertainty and of state evolution model uncertainty, respectively. Here, we employ the MMS to take account of possible appearance of outliers in the measurement, in the context of nonlinear state filtering. Three candidate measurement models including one Gaussian and two Student's distribution models are employed together to represent the measurement. By virtue of a model averaging mechanism, the effect of each model is dynamically adjusted according to the posterior distribution of each model, which is updated sequentially as we observe more data. The method allows the heavier tailed Student's t models to dominate the Gaussian model when the outliers arrive. This is done autonomously and dynamically within the PF algorithmic framework. The validity of our method is evaluated by illustrative simulations.
\section{Particle Filter}
In this Section, we give a succinct description for the PF algorithm. For more details, readers can refer to \cite{arulampalam2002tutorial,smith2013sequential,doucet2000sequential}.
Let us first consider a state space model:
\begin{eqnarray}
x_k&=&f(x_{k-1})+u_k\\
y_k&=&h(x_k)+n_k,
\end{eqnarray}
where $x_k\in\mathbb{R}^{d_x}$ and $y_k\in\mathbb{R}^{d_y}$ denote the target state vector and the measurement at the $k$th time step, respectively; $d_x$ and $d_y$ denote the corresponding dimensions. $f$ and $h$ denote the nonlinear state evolution function and measurement function, respectively. $u_k$ and $n_k$ represent independent identically distributed (i.i.d.) process and measurement noise sequence, respectively.
The probability density functions (pdfs) of $u_k$ and $n_k$, which are usually specified by the modeler, defines the state transition prior density $p(x_k|x_{k-1})$ and the likelihood function $p(y_k|x_k)$, respectively.

The Bayesian state filtering problem consists of computing the a posteriori pdf of $x_k$ given $y_{0:k}=\{y_i\}_{i=0}^k$, denoted by $p(x_k|y_{0:k})$ (or in short $p_{k|k}$). Recursive solutions are more preferable to batch mode methods; and, indeed $p_{k|k}$ can be computed from  $p_{k-1|k-1}$ recursively as follows
\begin{equation}\label{eqn:filter}
p_{k|k}=\frac{p(y_k|x_k)\int p(x_k|x_{k-1})p_{k-1|k-1}dx_{k-1}}{p(y_k|y_{0:k-1})}.
\end{equation}

The PF algorithm is an approximate solution to Eqn.(3) based on the sequential application of importance sampling (IS) techniques. Suppose that, at time step $k-1$, we have a discrete approximation of $p(x_{0:k-1}|y_{0:k-1})$ given by a set of weighted samples $\{x_{0:k-1}^i,\omega_{k-1}^i\}_{i=1}^N$, in which $x_{0:k-1}^i\sim q(x_{0:k-1}|y_{0:k-1})$,
$\omega_{k-1}^i\propto p(x_{0:k-1}|y_{0:k-1})/q(x_{0:k-1}|y_{0:k-1})$, $\sum_{i=1}^N\omega_{k-1}^i=1$. At time $k$, the $i$th trajectory is first extended by a particle $\hat{x}_k^i$ sampled from an importance distribution $q(x_k|x_{k-1},y_{0:k})$ and then weighted by
\begin{equation}
\omega_k^i\propto\omega_{k-1}^ip(\hat{x}_k^i|x_{k-1}^i)p(y_k|\hat{x}_k^i)/q(\hat{x}_k^i|x_{k-1}^i,y_{0:k}).
\end{equation}
It is well known that the above algorithm suffers from particle degeneracy when it is applied sequentially \cite{doucet2000sequential}. Precisely, after some iterations only few particles have a non null positive weight. A common practice to get around of this problem is to use after the weighting step
a resampling step meant to discard the particles with low weights and duplicate those with high
weights. Several resampling techniques have been proposed, see e.g. \cite{douc2005comparison,Li2015Resampling,Hol2006on}. A main scheme for an iteration of the PF algorithm can be summarized as follows. Starting from $\{x_{k-1}^i,\omega_{k-1}^i\}_{i=1}^N$:
\begin{itemize}
\item Sampling step. Sample $\hat{x}_k^i\sim q(x_k|x_{k-1}^i,y_{0:k})$, for all $i$, $1\leq i\leq N$;
\item Weighting step. Set $\omega_k^i$ using Eqn.(4) for all $i$, $1\leq i\leq N$, and normalize these weights to guarantee that $\sum_{i=1}^N\omega_k^i=1$;
\item Resampling step. Sample $x_k^i\sim\sum_{j=1}^N\omega_k^j\delta_{\hat{x}_k^j}$, set $\omega_k^i=1/N$, for all $i$, $1\leq i\leq N$. $\delta_{x}$ denotes the Dirac-delta function located at $x$.
\end{itemize}
\section{The Proposed Robust Particle Filter}
Here the measurement noise $n_k$ in Eqn.(2) is modeled by $M$ candidate models together. Let $\mathcal{H}_k=m$ denote the event that the $m$th model, $\mathcal{M}_m$, is the best one for use at time $k$.
Based on the Bayesian model averaging strategy \cite{hoeting1999bayesian,raftery1997bayesian,wintle2003use}, the posterior pdf under this multiple model setting is calculated as follows
\begin{equation}
p_{k|k}=\sum_{m=1}^Mp_{m,k|k}\pi_{m,k|k},
\end{equation}
where $p_{m,k|k}\triangleq p(x_k|\mathcal{H}_k=m,y_{0:k})$ and $\pi_{m,k|k}\triangleq p(\mathcal{H}_{k}=m|y_{0:k})$.
A recursive solution to compute Eqn.(5) is of particular interest here. Assume that at time $k-1$, we have at hand $\pi_{m,k-1|k-1}$, for all $m$, $1\leq m\leq M$, and a weighted sample set, $\{x_{0:k-1}^i,\omega_{k-1}^i\}$, which satisfies
\begin{equation}\label{eqn:particle_approx_t-1}
p_{k-1|k-1}\simeq\sum_{i=1}^N\omega_{k-1}^i\delta_{x_{k-1}^i}.
\end{equation}
At time $k$, the $i$th trajectory is first extended by a particle $\hat{x}_k^i$
sampled from an importance distribution $q(x_k|x_{k-1},y_{0:k})$ and then weighted by a weight
\begin{equation}\label{eqn:omega_mm}
\omega_{m,k}^i\propto\omega_{k-1}^ip(\hat{x}_k^i|x_{k-1}^i)p_m(y_k|\hat{x}_k^i)/q(\hat{x}_k^i|x_{k-1}^i,y_{0:k}),
\end{equation}
under the hypothesis $\mathcal{H}_{k}=m$, where $p_m(y_k|x_k)$ denotes the likelihood function associated with $\mathcal{M}_m$.
According to the IS principle, we have
\begin{equation}
p_{m,k|k}\simeq\sum_{i=1}^N\omega_{m,k}^i\delta_{\hat{x}_{k}^i}.
\end{equation}
Now let us consider, given $\pi_{m,k-1|k-1}$, how to derive out $\pi_{m,k|k}$.
First we specify a model transition process in term of forgetting \cite{liu2011instantaneous}, in order to predict the model indicator $\mathcal{H}$.
Let $\alpha$, $0<\alpha<1$, denote the forgetting factor. Given $\pi_{m,k-1|k-1}$, we have
\begin{equation}\label{model_pred_forget}
\pi_{m,k|k-1}=\frac{\pi_{m,k-1|k-1}^{\alpha}}{\sum_{m=1}^M\pi_{m,k-1|k-1}^{\alpha}},
\end{equation}
where $\pi_{m,k|k-1}\triangleq p(\mathcal{H}_{k}=m|y_{0:k-1})$. Then, employing Bayes' rule we have
\begin{equation}\label{posterior_model_indicator}
\pi_{m,k|k}=\frac{\pi_{m,k|k-1}p_m(y_k|y_{0:k-1})}{\sum_{m=1}^M\pi_{m,k|k-1}p_m(y_k|y_{0:k-1})},
\end{equation}
where $p_m(y_k|y_{0:k-1})$ is the marginal likelihood of
$\mathcal{M}_m$ at time $k$, defined to be
\begin{equation}\label{marginal_lik}
p_m(y_k|y_{0:k-1})=\int p_m(y_k|x_k)p(x_k|y_{0:k-1})dx_k.
\end{equation}
Here the state transition prior is adopted as the importance distribution, namely $q(x_k|x_{k-1},y_{0:k})=p(x_k|x_{k-1})$. Accordingly we have $p(x_k|y_{0:k-1})\simeq\sum_{i=1}^N\omega_{k-1}^i\delta_{\hat{x}_k^i}$. Then the integral in Eqn.(\ref{marginal_lik}) can be approximated as follows
\begin{equation}\label{eqn:particle_marignal_lik}
p_m(y_k|y_{0:k-1})\simeq\sum_{i=1}^N \omega_{k-1}^ip_m(y_k|\hat{x}_k^i).
\end{equation}
To summarize, one iteration of the proposed robust particle filter (RPF) is as follows. Starting from $\{x_{k-1}^i,\omega_{k-1}^i\}_{i=1}^N$ and $\pi_{m,k-1|k-1}$, for all $m$, $1\leq m\leq M$:
\begin{itemize}
\item Sampling step. Sample $\hat{x}_k^i\sim q(x_k|x_{k-1}^i,y_{0:k})$, for all $i$, $1\leq i\leq N$;
\item Weighting step. Set $\omega_{m,k}^i$ using Eqn.(\ref{eqn:omega_mm}) for all $i$, $1\leq i\leq N$, and normalize these weights to guarantee that $\sum_{i=1}^N\omega_{m,k}^i=1$, for all $m$, $1\leq m\leq M$;
\item Model Averaging step. Compute the posterior pdf of the model indicator, $\pi_{m,k|k}$, using Eqns.(9)-(12).
\item Resampling step. Sample $x_k^i\sim\sum_{j=1}^N\omega_k^j\delta_{\hat{x}_k^j}$, in which $\omega_k^j=\sum_{m=1}^M\pi_{m,k|k}\omega_{m,k}^j$; Set $\omega_k^i=1/N$, for all $i$, $1\leq i\leq N$.
\end{itemize}
The presented RPF uses three measurement noise models, including two Student's t distribution models and one Gaussian model, based on which the likelihood functions $p_m(y_k|x_k)$, $m=1,2,3$, are defined. All distribution models are zero mean with the a fixed covariance $\Sigma$. The involved two Student's t models discriminate with each other by the parameter, degrees of freedom (DoF). The DoF values under use are 3 and 50, corresponding to an extremely and an intermediate-level heavier tailed distributions, respectively.
Suppose that $x$ is a $d$ dimensional random variable that follows the multivariate Student's $t$ distribution,
denoted by $\mathcal{S}(\cdot|\mu,\Sigma,v)$, where $\mu$ denotes the mean and $v\in(0,\infty]$ is the DoF. Then the density function of $x$ is:
\begin{equation}\label{def_t}
\mathcal{S}(x|\mu,\Sigma,v)=\frac{\Gamma(\frac{v+d}{2})|\Sigma|^{-0.5}}{(\pi v)^{0.5d}\Gamma(\frac{v}{2})\{1+M_d(x,\mu,\Sigma)/v\}^{0.5(v+d)}},
\end{equation}
where
\begin{equation}
M_d(x,\mu,\Sigma)=(x-\mu)^T \Sigma^{-1}(x-\mu)
\end{equation}
denotes the Mahalanobis squared distance from $x$ to $\mu$ with respect to $\Sigma$, $A^{-1}$ denotes the inverse of $A$ and $\Gamma(\cdot)$ denotes the gamma function.
\section{Simulations}
We evaluated the validity of the proposed algorithm using the time-series experiment presented in \cite{van2000the}. The time-series is generated by the following
state evolution model
\begin{equation}
x_{k+1}=1+\sin(0.04\pi\times(k+1))+0.5x_k+u_k,
\end{equation}
where $u_k$ is a Gamma(3,2) random variable modeling the process noise. The observation model is
\begin{equation}
y_k=\left\{\begin{array}{ll}
0.2x_k^2+n_k,\quad\quad\quad k\leq30 \\
0.2x_k-2+n_k,\quad\, k>30 \end{array} \right.
\end{equation}
The goal is to estimate the underlying clean state sequence $x_k$ online based on the noisy observations, $y_k$, for $k=1,\ldots,60$.
\subsection{Case I: filtering without the presence of outliers}
First we considered the case without outliers. In this case the measurement noise, $n_k$, was drawn from a zero-mean Gaussian distribution.
A few different PF algorithms were used for performance comparison. The experiment was repeated 30 times with random re-initialization for each run. All of the PFs used 200 particles and the residual resampling \cite{Hol2006on}. The forgetting factor of RPF $\alpha$ takes a value of 0.9. The performance of the different filters is summarized in Table 1, wherein the EKPF and UPF denote the PFs which employ the extended Kalman filter and the unscented Kalman filter to generate the importance distribution, respectively. The table shows execution time (in seconds), the means and variances of the mean-square-error (MSE) of the state estimates. All the reported computing times are based on a computer equipped with an Intel i5-3210M 2.50 GHz processor with one core. They do not involve any parallel processing. The result show that the proposed RPF is more accurate than the other competitor algorithms in the sense of MSE, with less execution time than the UPFs.
\begin{table}\centering\small
\begin{tabular}{c||c||c|c}
\hline %
Algorithm & Time & \multicolumn{2}{c}{MSE} \\
& &mean&var \\\hline
PF: Generic& 1.561 &0.350&0.056 \\\hline
PF: MCMC move step& 3.275&0.371&0.047 \\\hline
EKPF&2.958&0.280&0.015\\\hline
EKPF: MCMC move step&7.033&0.278&0.013\\\hline
UPF&9.095&0.055&0.008\\\hline
UPF: MCMC move step&19.735&0.052&0.008\\\hline
the proposed RPF&5.509&0.018&0.0001\\\hline
\end{tabular}
\label{Table:convergence values}
\caption{Execution time (in seconds), Mean and variance of the MSE calculated over 30 independent runs for Case I.}
\end{table}
\subsection{Case II: filtering with the presence of outliers}
Next we designed a simulation case that involves outliers. The setting for the experiment time series was the same as Case I, except that several measurements at some time steps are replaced by outliers. The time steps associated with the presence of outliers are $k=7,8,9,20,37,38,39,50$. For typical measurements, their associated measurement noise was drawn from a zero-mean Gaussian distribution the same as for Case I. For outliers, the item $n_k$ in Eqn.(16) was drawn randomly from a uniform distribution between 40 and 50. All the considered algorithms were set to be blind to the above information on the outliers. The other settings for the experiment were the same as for Case I. The performance of the different filters is summarized in Table 2, which shows that the presented RPF method provides the most accurate online state estimation. For this case, the EKPFs and UPFs perform much worse than the other filters. We argue that it is due to the fact that both EKPFs and UPFs utilize the measurement information to build up the importance distribution, while, they will improperly take outliers as regular measurements upon the arrival of outliers and thus make the resulting importance distribution inefficient and misleading.
\begin{table}\centering\small
\begin{tabular}{c||c|c}
\hline %
Algorithm & \multicolumn{2}{c}{MSE}\\
&mean&var\\\hline
PF: Generic&0.533&0.040\\\hline
PF: MCMC move step&0.523&0.039\\\hline
EKPF&22.663&0.343\\\hline
EKPF: MCMC move step&22.668&0.358\\\hline
UPF&19.804&0.289\\\hline
UPF: MCMC move step&19.808&0.274\\\hline
the proposed RPF&0.357&0.010\\\hline
\end{tabular}
\label{Table:convergence values}
\caption{Mean and variance of the MSE calculated over 30 independent runs for Case II.}
\end{table}
\subsection{Further evaluations of RPF}
First we evaluated the sensitivity of the RPF's performance with respect to the forgetting factor $\alpha$.
We considered $\alpha$ values 0.1, 0.3, 0.5, 0.7 and 0.9. For each value, we ran the RPF algorithm 30 times for both Case I and II and calculated the corresponding mean of MSE. The result is depicted in Fig.\ref{fig:mse_alpha}, which shows that the performance of the presented RPF algorithm is not very sensitive to the selected values of $\alpha$, for both Case I and II.
\begin{figure}[t]
\centering
\includegraphics[width=3.3in,height=2.1in]{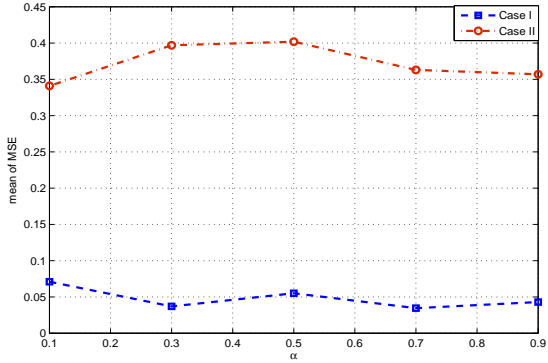}
\caption{Mean of the MSE calculated over 30 independent runs, in case of different $\alpha$ values, for both Case I and II.}\label{fig:mse_alpha}
\end{figure}

Next we fixed the value of $\alpha$ to be 0.1, and recorded the averaged posterior probability of each candidate measurement model at each time step over 30 times of experiments for both Case I and Case II. The result is plotted in Fig.\ref{fig:model_prob}. It is shown that, for both cases, the Student's t ($v$=3) model always dominates the other models.
The curves have no obvious patterns for Case I; but have an obvious pattern for Case II, that is, the posterior probability of the Student's t ($v$=3) model increases along with the appearance of outliers. Specifically, once an outlier appears (corresponding to time steps $k=7,8,9,20,37,38,39,50$), the posterior probability of the Student's t ($v$=3) model increases suddenly to a value close to 1; meanwhile, the posterior probabilities of the other two models decrease to 0 correspondingly.
\begin{figure}[t]
\centering
\includegraphics[width=3.3in,height=2.1in]{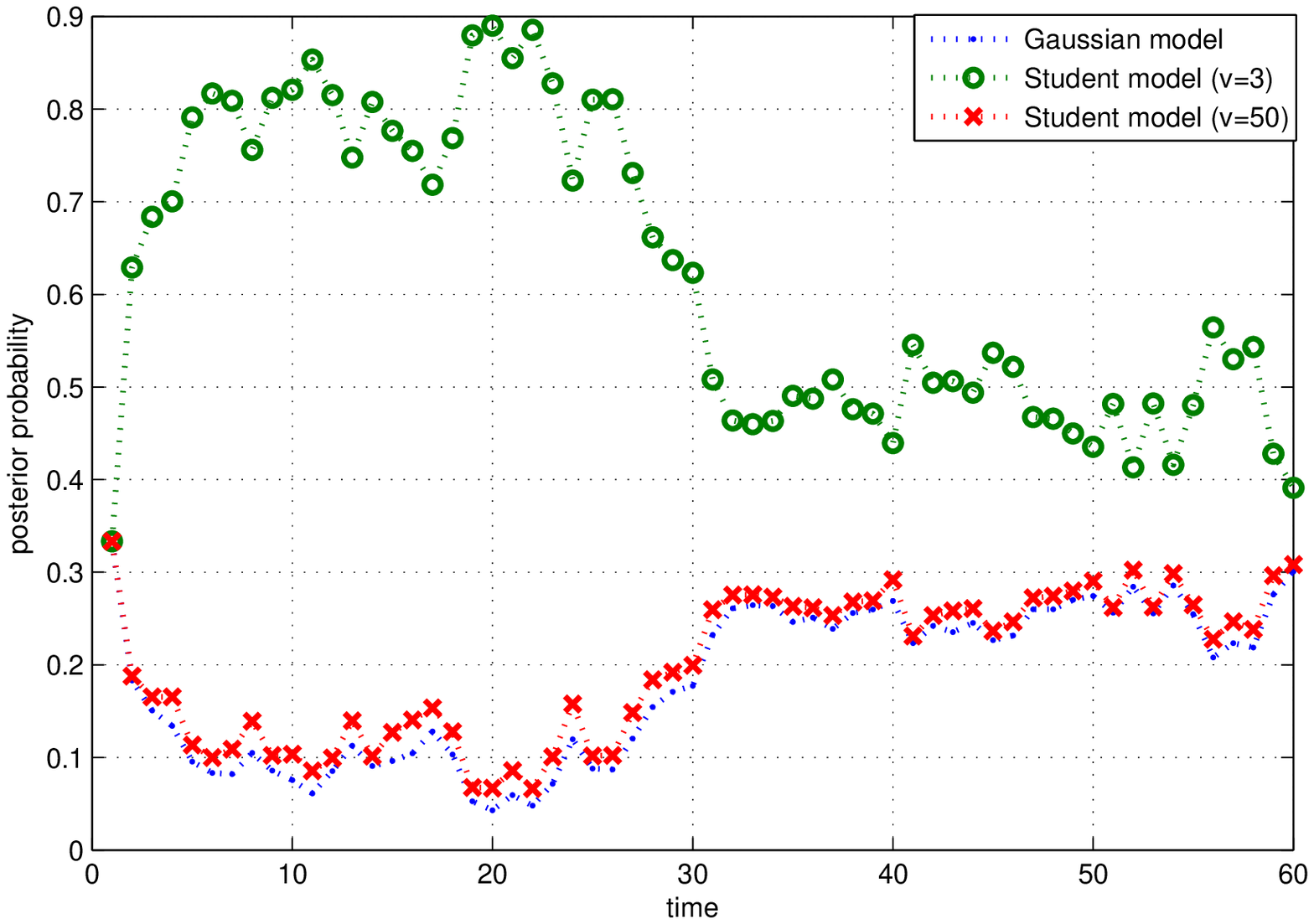}\\
\includegraphics[width=3.3in,height=2.1in]{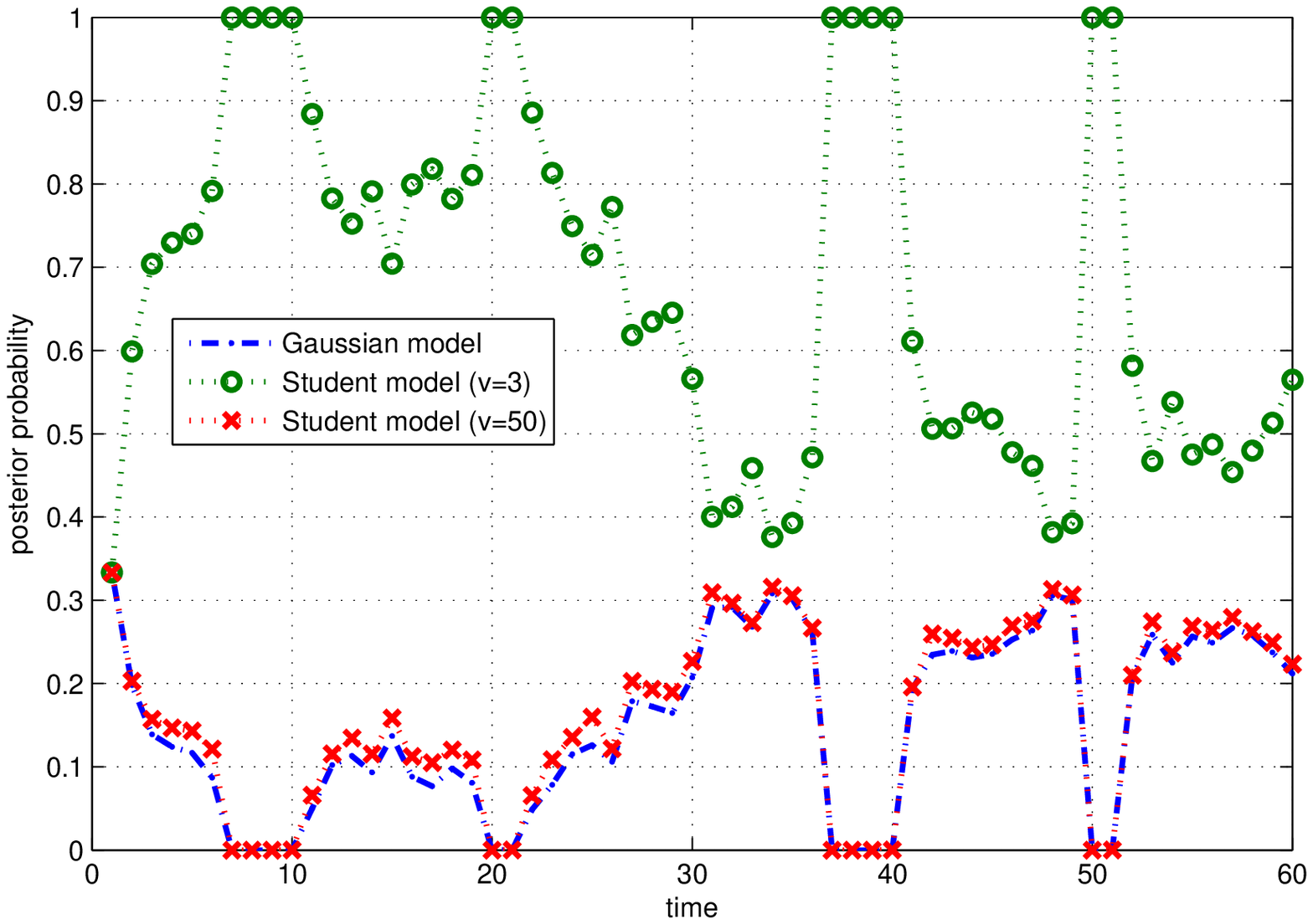}
\caption{Averaged posterior probability of candidate models outputted by the proposed RPF method. The top and bottom sub-figures correspond to Case I and II, respectively.}\label{fig:model_prob}
\end{figure}

Observing that the posterior probability of the Student's t ($v$=3) model is always much bigger than the others, we wondered if a PF algorithm which only employs the Student's t ($v$=3) model can produce the similar performance as the presented RPF algorithm. We set $\alpha=0.9$ and repeated the experiment of running the single Student's t ($v$=3) model based PF in the same way as described before for Case I and II. The resulting mean and variance of the MSE are presented in Table 3. In contrast with the performance of RPF as presented in Tables 1 and 2, we see that the single model based PF performs similarly as the presented RPF for case II, while, it loses in terms of MSE against RPF for case I.
\begin{table}[htbp]\centering\small
\begin{tabular}{c||c|c}
\hline %
 & \multicolumn{2}{c}{MSE}\\
&mean&var\\\hline
Case I &0.060&0.012\\\hline
Case II&0.366&0.006\\\hline
\end{tabular}
\label{Table:convergence PF_t}
\caption{Mean and variance of the MSE calculated over 30 independent runs of the Student's t (v=3) model based PF.}
\end{table}
\section{Conclusions}
In this paper, we propose a multi-model based PF method, which is robust against the presence of outliers in the measurements.
In the proposed RPF method, the heavier-tailed Student's t models are employed together with the conventionally used Gaussian model to represent the measurement noise. A Bayesian model averaging strategy is adopted to handle the issue of model uncertainty. It is shown that the proposed method is able to dynamically adjust the effect of each candidate model in an automatic and theoretically sound manner. The validity of this method is evaluated via illustrative simulations. Empirical results show that the RPF method performs strikingly better than several existent PFs for all cases under consideration. Future work lies in adapting the presented RPF algorithm to deal with real-life problems, e.g., sonar/radar target tracking \cite{liu2010multi}, in which the behaviors of the outliers may be more complex.

\bibliographystyle{IEEEbib}
\bibliography{mybibfile}
\end{twocolumn}
\end{document}